\documentclass[twocolumn,amsmath,amssymb,amsthm]{revtex4-2}
\pdfoutput=1
\usepackage{latexsym}
\usepackage{soul,xcolor}
\usepackage{times}
\usepackage[T1]{fontenc}
\usepackage[utf8]{inputenc}
\usepackage{url}
\usepackage{hyperref}
\usepackage{graphicx}
\usepackage{epsfig}
\usepackage{mathptmx}
\usepackage{bm}
\usepackage{array}
\setstcolor{red}

\usepackage{draftcopy}

\begin{document}

\title{Self-organised dynamics beyond scaling of avalanches: Cyclic stress fluctuations in critical sandpiles}
\author{Bosiljka Tadi\'c$^1$, Alexander Shapoval$^2$, and Mikhail Shnirman$^3$}
\affiliation{$^1$Department of Theoretical Physics, Jo\v zef Stefan Institute,
Jamova 39, Ljubljana, Slovenia}
\affiliation{$^2$Department Mathematics and Computer Science, University of Lodz, Banacha 22, \L\'od\.z 90-238, Poland}
\affiliation{$^3$Institute of Earthquake Prediction Theory and Mathematical Geophysics, RAS, Profsoyuznaya 84/32, 117997 Moscow, Russia}
\vspace*{3mm}
\date{22.03.24}

\begin{abstract}
Recognising changes in collective dynamics in complex systems is essential for predicting potential events and their development.  Possessing intrinsic attractors with laws associated with scale invariance, self-organised critical dynamics represent a suitable example for quantitatively studying changes in collective behaviour. We consider two prototypal models of self-organised criticality, the sandpile automata with deterministic (Bak-Tang-Wiesenfeld) and probabilistic (Manna model) dynamical rules, focusing on the nature of stress fluctuations induced by driving---adding grains during the avalanche propagation,   and dissipation through avalanches that hit the system boundary.  Our analysis of stress evolution time series reveals robust cycles modulated by collective fluctuations with dissipative avalanches.   These modulated cycles are multifractal within a broad range of time scales. Features of the associated singularity spectra capture the differences in the dynamic rules behind the self-organised critical states and their response to the increased driving rate, altering the process stochasticity and causing a loss of avalanche scaling.  In the related sequences of outflow current, the first return  distributions are found to follow modified laws that describe different pathways to the gradual loss of cooperative behaviour in these two models. The spontaneous appearance of cycles is another characteristic of self-organised criticality. It can also help identify the prominence of self-organisational phenomenology in an empirical time series when underlying interactions and driving modes remain hidden.
\end{abstract}

\maketitle

\section{Introduction\label{sec:intro}}
The cooperative behaviour of interacting units is at the heart of emergent
features in many complex systems \cite{ComplexSystems_Estrada2023}; therefore, understanding changes in collective dynamics is vital for predicting their evolution.
Large interacting nonlinear systems, driven out of equilibrium, often
exhibit cyclical trends in the temporal evolution of a relevant
quantity (see recent study \cite{we_cycles_emo23} and references
there). The appearance of cycles can be visualised as a temporal
accumulation of 'energy' and then its release through collective
dynamics involving many units. Imperfect (modulated) cycles were observed everywhere,  from geophysical and solar irradiance cycles \cite{Cyc_solar2010}, which impact the climate and life on the Earth, to physics laboratory systems driven by an external magnetic field
\cite{mijatovic2022tuneable}, traffic on networks
\cite{Cyc_Traffic_dynamics2022}, and urban growth \cite{Cyc_urbangrowth2021}. Furthermore, the interplay of bio-social processes
\cite{Cyc_Epi2014recurrent,Cyc_Epi2022we} induces complex epidemic cycles, and social activity driven by the circadian day-night cycle crucially affects social dynamics
\cite{we_cycles_emo23}. These modulated cycles exhibit higher
harmonics that can be described by multifractal analysis
\cite{Cyc_gaoJSTA2009algorithm,Cyc_Traffic_dynamics2022,we_cycles_emo23}. In general, the mechanisms of their appearance in different systems still need to be better understood.
On the other hand, some nonlinear dynamical systems, which are
\textit{repeatedly} driven by external forces observing the time scale separation, can evolve towards attractors with critical dynamics. Long-range correlations and scaling behaviour of avalanches characterise these self-organised critical (SOC) states; see recent
review \cite{SOC_review2021} and references therein. They represent a specific type of collective dynamics with scale invariance, allowing
us to decipher a few (out of potentially many) parameters that
determine the universal critical behaviours and, thus, study
collective dynamics in greater detail. In this work, we examine
prototypical sandpile automata models and reveal that cycles emerge spontaneously as another prominent feature of SOC dynamics. Monitoring their modulation can be a good measure of changes in collective behaviours.

The Abelian sandpile automata and related models
\cite{dhar1999AbelianSPA},  as the paradigm of SOC, provides
theoretical ground to study complex features of self-organised
critical states: slow driving, avalanches, the relevance of time-scale
separation and dissipation at the system's boundaries. Well-studied
models Bak-Tang-Wiesenfeld (BTW) \cite{SPA_btw} with deterministic and
Manna model (MM) \cite{SPA_manna1991two} with probabilistic distribution rules differ in the finite-size scaling properties of the avalanches \cite{tebaldi1999multifractal}, even though the avalanche scaling 
exponents are numerically similar. Moreover, the sandpile automata models are also the
focus of the studies on predictability \cite{shapoval2021predictability,SOC_prediction2022Frontiers} and
information complexity
\cite{formenti2020hard,ComputCompl_freesingSPA2021},
motivated by the fact that a critical state possesses an efficient way
to store information. Universality of SOC  can be affected by the
geometry of underlying space
\cite{SOC_we_dynamics2021,kalinin2018self,SPA_BethelattDD1990,bhaumik2020critical},
randomness \cite{tadic1998disorder} and coupled environment dynamics
\cite{antonov2021effects,antonovSOC2023,antonov2023symetry}, as well
as altered probabilistic toppling conditions
\cite{we_pSPAdeepak1997,we_granularElowPRE1998,SOC_facilitatedTopplingPRL2016}, activation beyond toppling dynamics \cite{SOC_activeSPAentropy2023}
and autonomously adapting  \cite{SOC_AdaptingSPAmm_Gross2020} sandpiles. With the finite driving rates \cite{we_avalQueusPRE2000}, grains are added \textit{during the
avalanche propagation}, which may locally alter the strict toppling
rules and trigger additional event sequences. Consequently,  changed
scaling properties of the avalanches
\cite{BHNavalanches_Drivingrate2021} and  possible loss of scaling may
occur, depending on the system size and dissipation, when the driving
rate exceeds a specific critical value
\cite{SOC_drivingrates2010}. Thus, studying the time-dependent
properties \cite{SOC_timedepFrontiers2021,SOC_timedepSasha2022,SOC_oscillNeuronsSciRep2019} are necessary to reveal salient features of self-organised dynamics beyond the scaling of avalanches.

Many complex systems show signatures of SOC
\cite{SOC_Jensen,SOC_Epi2000,SOC_EarthSystems2002,aschwanden2013self,markovic2014power,SOC_we_dynamics2021},
where it is recognised as a  'blueprint for complexity' \cite{SOC_weHBNets2019}, mechanisms providing robustness in steady
states and functional properties
\cite{bak1995complexity,wolf2018physical}, or  a 'trade-off between
cooperation and competition' \cite{SOC_econo2021review}.
SOC is evidenced by numerical methods \cite{mcateer201625} from
available empirical data, e.g., time series of a relevant quantity,
for example, brain activity \cite{gros2021devil}, epidemic processes
\cite{SOC_Epi2014qexpdengue} and online social cooperation \cite{SOC_wePRE2017knowe}, or 
geophysical and solar activity \cite{smyth2019self} and rainfalls
\cite{SOC_rainfalls2015}. 
Similarly, properties of SOC are utilised to solve complex optimisation problems \cite{SOC_Optimization2018}, traffic congestion management \cite{SOC_trafficmanagement2023} and design robust functional systems
such as computer networks \cite{SOC_compnets2002Valverde}.  Most of
these systems operate under time-varying driving fields that change at a finite rate compared to a typical avalanche propagation time. 
Therefore, recognising the mechanisms of self-organisation from the structure of a time series of a relevant variable is critically important for identifying dynamical states in many complex systems where the interactions and driving modes are less apparent.

In this work, using two well-known sandpile automata with
deterministic (BTW) and probabilistic (MM) toppling rules, we study the sandpile dynamics at adiabatic driving with a perfect
time-scale separation and at several finite driving rates where
additional grains are dropped \textit{during the avalanche propagation}.
We reveal the emergence of robust cyclical trends at a slow time scale in
the temporal evolution of stress, defined as the number of grains in
the sandpile divided by lattice area. 
 Dissipative avalanches modulate these cycles in a broad range of time scales, characterised by the appropriate multifractal measures. The shapes of the respective singularity spectra correlate with
the sandpile automaton rules and their characteristic response to changed driving rates. In conjunction with the altered sequences of
outflow currents, these multifractal features suggest fundamental differences in the underlying self-organised dynamics beyond the scaling of avalanches. 
In a more general context, these measures can be helpful to identify
changes in the collective dynamics, and may be beneficial in studies of
the empirical data from different complex systems with expected
self-organised behaviours.

\section{Simulations and results\label{sec:results}}

\subsection{Definition of the models at finite driving rates}\label{sec:mod}

We consider a modification of the Bak-Tang-Wiesenfeld (BTW)
\cite{SPA_btw} and Manna model (MM) \cite{SPA_manna1991two}  sandpiles
 on the square lattice with open boundaries, where
the system is driven by the inflow of grains (the unit of stress)
at random locations but  at a fixed rate \textit{during the avalanche propagation}.
Here is the specific for the \emph{BTW model}.
Let $\mathcal{L} = \{(i,j)\}_{i,j=1}^L$ denote the $L \times L$ lattice.
Then integers $h_{ij}$ interpreted as stress are assigned to each cell $(i,j) \in \mathcal{L}$. Their set $\mathcal{H} = \{h_{ij}\}_{i,j=1}^L$ is the configuration of stress.
As in the original model, the instability threshold $h_c$ for the stress
in the cell is introduced, where $h_c= h_{\text{BTW}}=4$, representing
the number of neighbors of any inner cell $(i,j)$, where
the cells sharing a side with $(i, j)$ form the set $\mathcal{N}(i,j)
= \mathcal{N}_{\text{BTW}}(i,j)$ of its neighbors are 
\begin{equation}
  \mathcal{N}_{\text{BTW}}(i,j) = \{(i',j') \in \mathcal{L}: |i'-i|+|j'-j| = 1 \}.
  \label{e:neighborBTW}
\end{equation}
The cells $(i,j)$ with $h_{ij} < h_c$ are called stable.
The configuration $\mathcal{H}$ is called stable if all cells are stable.
In contrast, the inequality $h_{ij} \ge h_c$ indicates that the cell $(i,j)$
is unstable, and the configuration is unstable if it consists of at least
one unstable cell.

The following automata update rules describe the dynamics.
Let $h_{ij}(t)$ denote the value of $h_{ij}$ at the beginning of the
time moment $t$.
Then a random cell $(i,j)$ is chosen and the corresponding $h_{ij}$
is increased by $1$
\begin{equation}
  \begin{aligned}
    & h_{ij}(t) = h_{ij} \longrightarrow h_{ij}+1,
    \\
    &\text{$(i,j)$ \texttt{is random from } $\{1,\ldots,L\} \times \{1,\ldots,L\}$}.
  \end{aligned}
  \label{e:add}
\end{equation}
If the resulting value of $h_{ij}$ is still less than $h_c$
then nothing more occurs at this time moment.
Otherwise, the violation of the stability condition triggers the avalanche,
formally defined in the following way.\\
\textbf{(i) Parallel updates of stress.}
The unstable cell transports $h_c = 4$ units of stress to 
the neighbours, a single unit to each:
\begin{gather}
  h_{ij} \longrightarrow h_{ij}-h_c
  \label{e:loss}
  \\
  h_{i'j'} \longrightarrow h_{i'j'}+1,
  \quad \forall (i',j') \in \mathcal{N}(i,j).
  \label{e:gain}
\end{gather}
This first update represents the first step in the avalanche propagation. It may result in the appearance of new unstable
cells. All these unstable cells also transport the stress to the neighbours
in line with equations~\eqref{e:loss}, \eqref{e:gain}, creating the second round of updates.
Let us say that these updates occur in parallel and, hence,
we have just defined the second round of parallel updates.
In general, if the round $k$ of the parallel updates ends up with 
an unstable configuration, then the transport from all existing unstable
cells in line with equations~\eqref{e:loss}, \eqref{e:gain} creates
the $(k+1)$th round of the parallel updates.
These updates occur as long as the configuration is
unstable. 
In the simulation, it is convenient to code the application of rules~\eqref{e:loss}, \eqref{e:gain}, describing a round of the parallel updates, one by one. Note that the order in which the unstable cells are chosen during any round of the parallel updates does not affect the resulting configuration.
\\
\textbf{(ii) Driving after every $R$th parallel update.} 
The adiabatic driving, i.e., the addition of new grain after an
avalanche stops, as defined in the original models, obeys a perfect
time scale separation between the driving and propagation of
avalanches.
Here, we introduce another algorithm parameter, $R > 1$,  implementing a finite driving rate.
If round $R$ of the parallel updates defined in (i) occur at the current
time moment $t$, then another grain   is added to a randomly chosen cell, equation~\eqref{e:add}.
This additional grain can increase the number of unstable cells
in the lattice. All unstable cells (if any exist) are processed
during the round $R+1$, and so on.
In the same manner, an additional unit of stress inflows to the random
site, as determined by~\eqref{e:add}, after each $R$th
round of the parallel updates, i.e., after round $pR$ for any natural $p$,
if this round has happened. Thus, the system is loaded at the rate  $1/R$ at the time scale associated with the avalanche propagation. \\
\textbf{(iii) Dissipation at the boundary.}
At the system's boundary, any non-corner cell $(i,j)$ has
$|\mathcal{N}(i,j)| = 3$ neighbors and the corner cell has
$|\mathcal{N}(i,j)| =2$ neighbors. Hence, when an unstable cell belongs to the boundary, the toppling rules~\eqref{e:loss}, \eqref{e:gain} lead to the dissipation of $h_c - |\mathcal{N}(i,j)|$ units of the stress out of the
lattice. 

Rules (i)--(iii) define entirely the parallel updates that continue
until the stable configuration $\mathcal{H} = \{h_{ij}\}_{i,j=1}^L$ is
attained. The values of the stress in this configuration are
associated with the next time moment $t+1$: $h_{ij}(t+1) = h_{ij}$,
$i,j=1,\ldots,L$.

The \emph{Manna model} with the additional $1/R$ driving rate at the fast time is defined similarly. We specify here the differences from the BTW case.
Specifically, the threshold $h_c=h_{\text{MM}} = 2$;
the set $\mathcal{N}_{\text{M}}(i,j)$ of neighbors of an inner cell $(i,j)$
consists of two neighbours that are sampled randomly from the set
$\mathcal{N}_{\text{BTW}}(i,j)$ as
\begin{multline}
  \mathcal{N}_{\text{MM}}(i,j) = 
  \{N_1, N_2: N_k,\, k=1,2,
  \\
  \text{\texttt{ is a random element from}
    $\mathcal{N}_{\text{BTW}}(i,j)$}\} \ .
  \label{e:neighborM}
\end{multline}
To apply the above definition~\eqref{e:neighborM} to the boundary cells
$(i,j)$ we extend $\mathcal{N}_{\text{BTW}}(i,j)$ by adding the empty
set $\emptyset$ such that the extended $\mathcal{N}_{\text{BTW}}(i,j)$
contains exactly four elements, including the empty set(s).
If the empty set is chosen once or twice at~\eqref{e:neighborM}, then
the corresponding set of neighbours consists of one element or becomes empty. In the latter case, two units of stress dissipate at the boundary
when the cell $(i,j)$ is updated.
We note that the determination of neighbours is required only to process updates. We sample the neighbours for a cell at the moment the  rules~\eqref{e:loss}, \eqref{e:gain} are applied, independently of earlier
samples related to this or other cells.
With the above changes with respect to the BTW model,
the dynamics is determined by the gradual loading, according to~\eqref{e:add}, in the slow time and the transport of stress
in the fast time governed by (i)--(iii). The quantities $h_{\text{MM}} = 2$
and $\mathcal{N}_{\text{M}}$ are used for $h_c$ and $\mathcal{N}$.
Thus, this framework defines two families of models depending on the
parameter $R$, which determines the rate $1/R$ of the driving at the
fast time scale. 
Note that large values of $R$ often exclude the driving during the avalanche evolution, and we end up with the original BTW and Manna sandpiles.

For each driving rate, starting with an empty lattice ($L=256$, all $h_{i,j}=0$), the
grains are gradually added, and the simulation is performed according
to the above-described model rules. After a specific transient time, the system reaches a stationary state, where  the average stress 
$h(t) = L^{-2}\sum_{i,j=1}^L h_{ij}(t)$ fluctuates  around a well-defined
level. We then start recording the data regarding $200\,000$ distinct avalanches.
We are sampling the evolution of the average stress,  the sequences of
the size and duration of all avalanches; we separately record the
sequences of  dissipative avalanches (i.~e., at least a single unit
of stress is dissipated during these avalanches),  and their outflow
current--the number of dissipated grains per avalanche. These quantities are further analysed in the following.

\subsection{Size and duration of avalanches at finite driving rates}\label{sec:sizefreq}
When an avalanche is triggered at an unstable site, its propagation
through the system is followed. The number of parallel updates before
reaching a stable configuration defines the duration $d$ of the
avalanche. Meanwhile, the size $s$ of an avalanche is the number of
cells that become unstable during the avalanche. Each unstable cell is
counted as many times as it is unstable. Then the probability density
function  $\hat{\varphi}(s)$ and $\hat{\phi}(d)$ of avalanche sizes
and durations are
derived from the catalogue of avalanche sequences. As stated above, we
consider  \textit{all} avalanches and, separately, the subset of \textit{dissipative} avalanches that hit the boundary.  

\begin{figure}[!htb]
\begin{tabular}{cc}                                                                                                       
\resizebox{16pc}{!}{\includegraphics{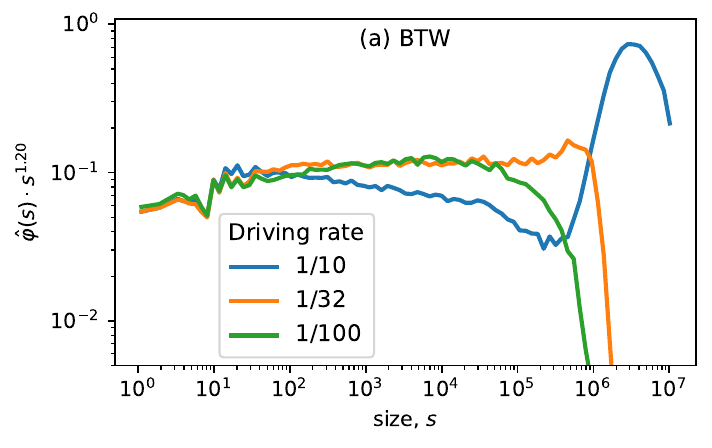}}\\
\resizebox{16pc}{!}{\includegraphics{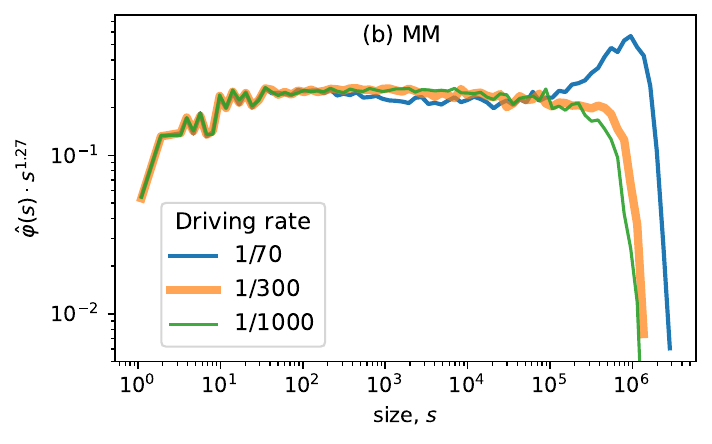}}\\
\end{tabular}
  \caption{\small The size-frequency relationship
    $\hat{\varphi}(s)s^{\tau_s}$ vs $s$ of all avalanches for
    different fast-time driving rates $1/R$ in the BTW (a) and MM (b)
    sandpiles and $\tau_s$ corresponding to adiabatic driving; see text.}
\label{f:sizefreq}
\end{figure}

The size-frequency relationship is likely the most known outcome
of the original BTW and MM sandpiles because of its power-law segment that extends through almost all scales up to the cut-off caused by the finite systems' size.
Fig.~\ref{f:sizefreq} illustrates the size-frequency relationship for 
our models built-in from the BTW and MM sandpile automata at
finite driving rates. Logarithmically binned data are shown. To
demonstrate how the avalanche scaling is lost at finite driving rates, 
we display $\hat{\varphi}(s)$ multiplied by an appropriate power
function of $s$, $s^{\tau_s}$, where the values of $\tau_s$,
$\tau_{s,\text{BTW}}=1.20$ and $\tau_{s,\text{MM}}= 1.27$ for the BTW
and Manna models respectively, are the exponents taken from earlier
studies at adiabatic driving~\cite{priezzhev1996formation} and~\cite{bhb96}.

\begin{figure}[!htb]
\begin{tabular}{cc}                                                                                                       
\resizebox{16pc}{!}{\includegraphics{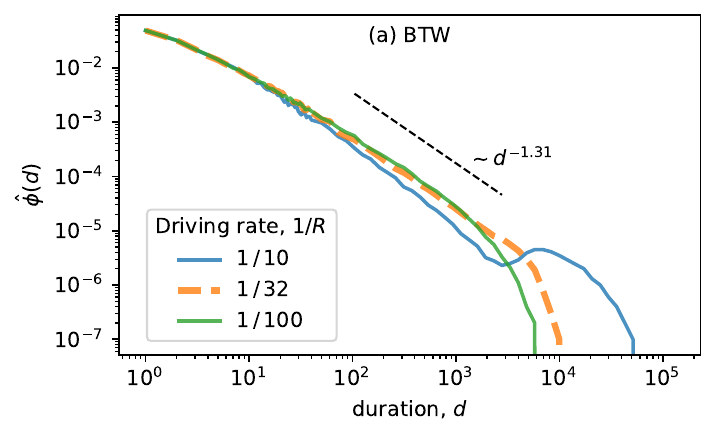}}\\
\resizebox{16pc}{!}{\includegraphics{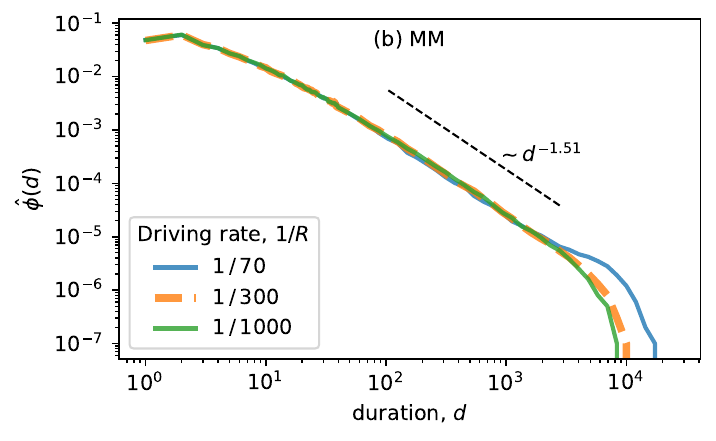}}\\
\end{tabular}
  \caption{\small The duration distribution $\hat{\phi}(d)$ vs
    duration $d$ of all avalanches at varied driving rates $1/R$ in the BTW (a) and MM (b) 
    sandpiles.}
\label{f:durfreq}
\end{figure}

\begin{figure}[!htb]
\begin{tabular}{cc}      
\resizebox{16pc}{!}{\includegraphics{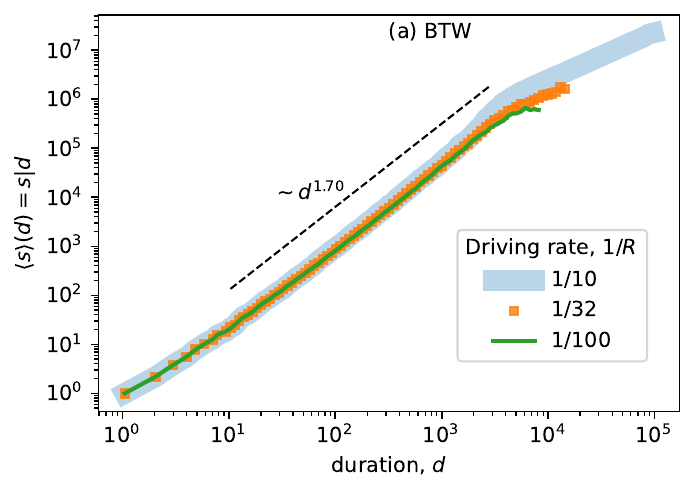}}\\
 \resizebox{16pc}{!}{\includegraphics{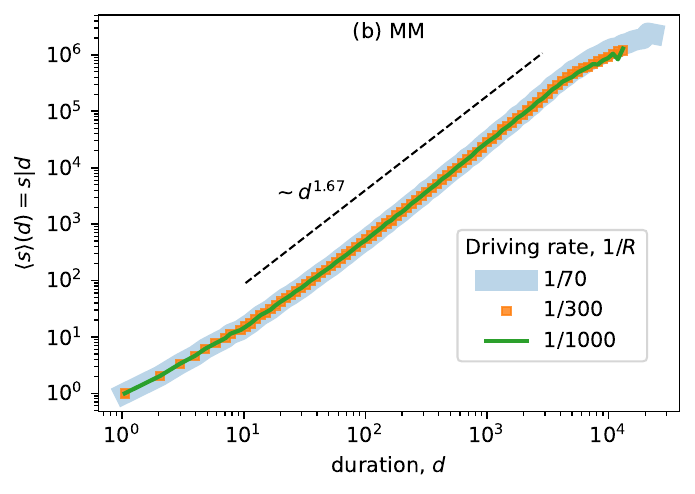}}\\                     \end{tabular}
  \caption{\small The average size of the avalanches with a given
    duration $<s>_d$ vs{.} duration $d$ for the BTW (a) and MM (b)
    sandpiles at different driving rates.}
\label{f:sizenorm-dur}
\end{figure}

The probability distribution functions $\hat{\varphi}(s)$ constructed
with different values of the parameter $R$ exhibit a critical change
for both families of models. 
In particular, as the fast-time driving rate increases, 
the power-law segment is preserved with slightly changed exponents until the driving exceeds a specific critical rate $\gtrsim
1/R^*$; then the scaling of avalanches is lost, and the sandpile
becomes frequently overloaded launching anomalously large
avalanches. The corresponding size and duration distributions exhibit
a bump at the right part, as shown in Figs.\ \ref{f:sizefreq}-\ref{f:durfreq}.
For the considered system size, we illustrate the estimated
$\hat{\varphi}(s)$ with
three values of $R$ correspond to low, (nearly) critical, and super-critical driving rates for both families in Fig.~\ref{f:sizefreq}.
Notably, the two families of the $\hat{\varphi}(s)$ curves also have differences in their dependence on $R$.
Specifically, in the BTW models, the power-law segment becomes longer
with the growth of the driving rate $1/R$ up to its critical threshold
in contrast to the MM, where it is practically unchanged until the
scaling is lost.
We recall that the principal difference between the original BTW
and Manna sandpiles (without additional driving) is in the distribution's tail. The latter is multifractal with the BTW sandpile but
admits the finite size scaling, applicable to the whole size-frequency
relationship~\cite{bhb96, tebaldi1999multifractal} in MM.
The prolongation of the power-law segment to the right with the growth of $1/R$
for the BTW models likely simplifies the distribution tails, possibly
altering its multifractality (this claim needs to be checked with an independent study).

In analogy to the size distribution, the growth in the driving rate $1/R$ up to a threshold value $1/R^*$
roughly conserves the shape of the duration distribution in the BTW
model; meanwhile, the driving rate beyond the critical value
restructures the shape of the $\hat{\phi}(d)$, as shown in  Fig.~\ref{f:durfreq}.
Note that, the power-law fragment with the expected exponent 
is not observed at slow drivning; perhaps, larger lattices are necessary.
However, for finite driving rates, the power-law segment with the
exponent $\tau_{d}\approx 1.31$ appears, whereas the bump precedes the fall at the tail. 
Similar observations  apply to MM as shown in  Fig.~\ref{f:durfreq};
The power-law segment  is observed in the regime below critical dring
rate, and the shape of the distribution is preserved.
Faster driving also creates a bump at the tail of the distributions. 
These properties of the avalanches at finite driving rates also manifest in the plots in Fig.\ \ref{f:sizenorm-dur}, where the average size of the avalanches of a given duration $<s>_d$ is shown against the duration $d$.
Interestingly, the size--duration scaling is preserved at finite
driving rates for the intermediate avalanches
Thus, both models with the driving rate at the fast time scale 
exhibits the power-law fragment suggesting the preserved scaling
relation between the size and duration of avalanches in the
intermediate range.

\begin{figure}[!htb]
\begin{tabular}{cc}                                                                                                 
\resizebox{16pc}{!}{\includegraphics{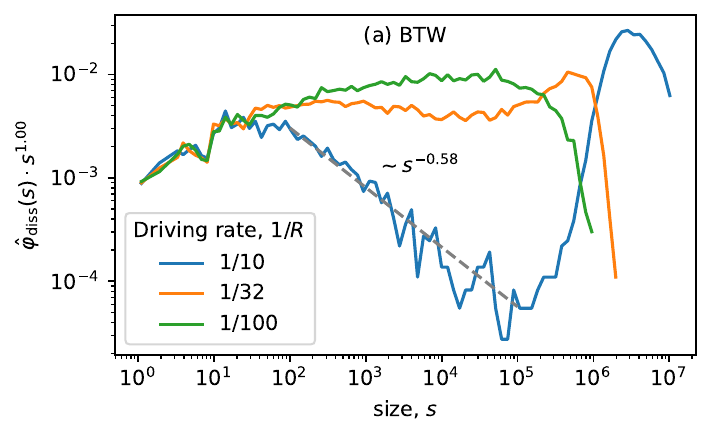}}\\
\resizebox{16pc}{!}{\includegraphics{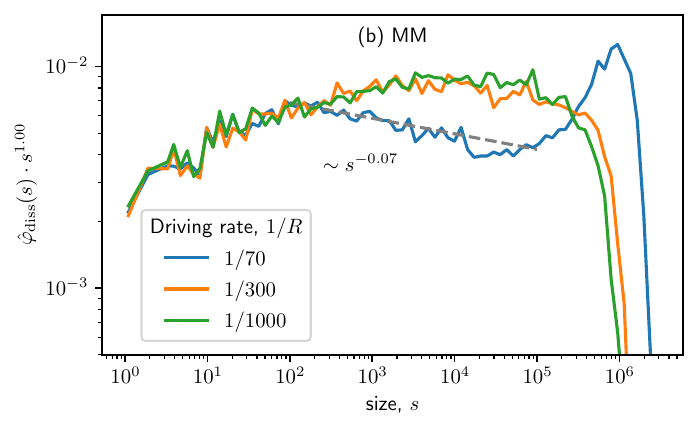}}\\
\end{tabular}
  \caption{\small The size-frequency relationship $
    \hat{\varphi}_{\text{diss}}(s)s^{1.00}$ vs. $s$ for the dissipative
    avalanches for the BTW (a) and MM (b) models for three
    driving rates $1/R$. 
The fit (in dashed grey) to the blue curve is
    shown (a) where it is computed and (b) in a broader interval than
    computed. 
}
\label{f:sfrdiss}
\end{figure}

The probability density functions$ \hat{\varphi}_{\text{diss}}(s)$
related to the subset of avalanches that hit the system's boundary,
known as dissipative avalanches, are shown in Fig.\ \ref{f:sfrdiss}.
Elucidating the importabce of dissipative avalanches in scaling of all
avalanches in BTW model,  paper~\cite{drossel2000scaling} indicated
the power-law distribution of dissipative avalanches in the original
BTW model at the adiabatic driving with a specific exponent
$7/8$; similarly, the exponent close to $1$ was derived
\cite{SOC_MMdissipavalanchesPRE2003} for MM.
We complement this general claim with the observation that the accuracy of  the power-law approximation is significantly smaller in the dissipative
avalanches at finite driving rates; compare the curves in Fig.~\ref{f:sizefreq}a.
and~\ref{f:sfrdiss}a).
The existence of the power-law segment for 
$\hat{\varphi}_{\text{diss}}$ with slow driving rates,  $R=100$, is
questionable.
However, the growth of the driving rate to $1/32$, which seems to be close to a critical value, purifies the power-law segment on $[2\cdot 10^2,
10^5]$, and the corresponding exponent is close to $1$.
Further growth in the driving rate keeps the power-law segment on the
same interval of sizes, but the exponent increases to $1.58$ (i.e., $0.58$
corresponding to the displayed axes).
With the Manna models, the general behaviour of the curves
$\hat{\varphi}_{\text{diss}}$ representing the size-frequency
relationship agrees with previous observations in Fig.\ \ref{f:sizefreq}b. The curves are similar if the driving rate is small (see orange and green curves in Fig.~\ref{f:sfrdiss}b).
For a larger driving rate $1/R$, a segment close to power-laws that admit approximations by the 
straight lines in the double logarithmic scale with a negative
slope (approximately, $1.07$)  appears; meanwhile, the initial segment
of all curves scales with the exponent which is less than $1$.
When the driving rate $1/R$ crosses a critical value,
the initial point of the second power-law segment is moved to the left
(see the blue line in Fig.~\ref{f:sfrdiss}b, where the fit is computed
with $s\in (87, 539)$ but prolonged further to the right).

\subsection{Stress fluctuations cycles \label{sec:cycles}}
\begin{figure*}[htb]
\begin{tabular}{cc} 
\resizebox{32pc}{!}{\includegraphics{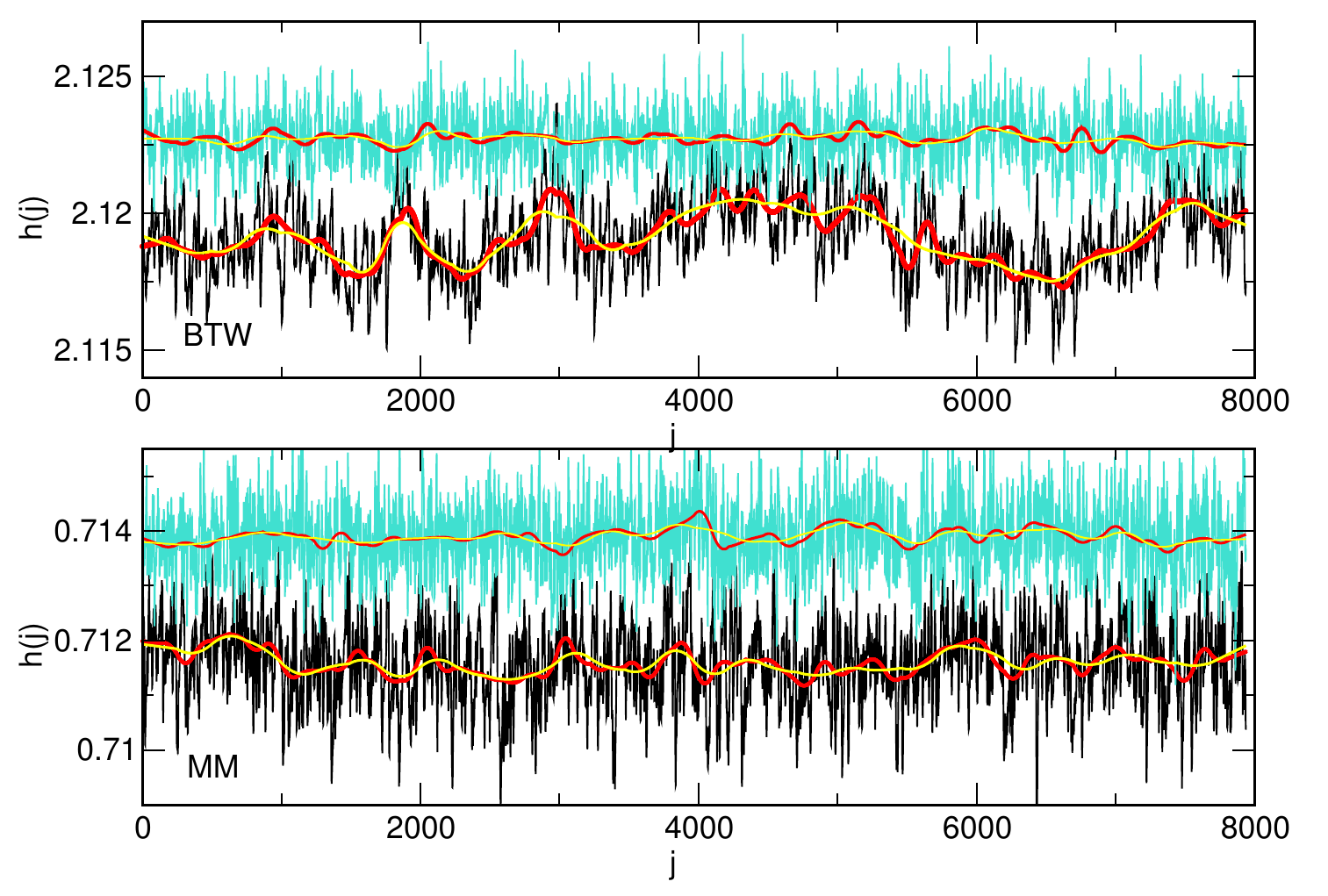}}
\end{tabular}
\caption{Stress $h(j)$ vs. time index $j$ in the BTW (top) and MM sandpiles (bottom)
  for two driving conditions: adiabatic slow driving (lower-black line)
  and a fast driving  at  $R=10$ in BTW and $R=70$ in MM (upper-cyan line;
  the line is shifted vertically for better vision in the case of
  MM). Thick red and yellow lines indicate the corresponding cycles identified by
  fixing $m=124$ and $m=248$, respectively; see text for details.   }
\label{fig:ts2x}
\end{figure*}

We recall that, apart from adding grains at a given rate,  the stress fluctuations are induced at the slow time scale by outflow at the system's boundary, which is carried by the dissipative avalanches. At this scale, the time series of stress fluctuations in the critical state for
the BTW and MM sandpiles depend on the driving rate $1/R$ at which
additional grains are dropped during the avalanche propagation, as  Fig.\ \ref{fig:ts2x} demonstrates;
because of slow varying stress, every 25th time step (distinct avalanche) is recorded,
indicated by the index $j$ along the time axis.
Here, we plot the case of adiabatic driving, i.e., dropping one grain
per avalanche, and the case where the average number of dropped grains
per avalanche is similar,
measured by the ratio of the average
duration at finite driving $<d>_R$ and the average duration $<d>_0$ at
adiabatic driving. Given the different propagation of avalanches in these two models, this ratio $<d>_R/<d>_0\approx 14 $, is reached at $R=10$ in BTW, and $<d>_R/<d>_0\approx 13$ for $R=70$ in MM.
Notably, these time series differ in the average stress level and the shape of the emergent irregular cycles. Moreover, the stress evolution at a finite driving rate in the BTW model, see the top
line in Fig.\ \ref{fig:ts2x}, is profoundly different from the case
where the driving is adiabatic, which preserves the original
deterministic toppling, see the black line in the same
panel. Meanwhile, the differences are visually more minor in the 
Manna model, which includes a certain degree of probabilistic toppling
even in the adiabatic driving mode. A striking feature of these SOC
states is the appearance of cyclical trends of stress evolution at
different time scales, as shown in Fig\ \ref{fig:ts2x} by solid red
and yellow lines.

We determine these trends by changing the parameter $m$ in  the local
adaptive detrending algorithm; this methodology is  introduced in 
\cite{Cyc_gaoJSTA2009algorithm} to analyse sunspot time series
associated with solar cycles and adapted to treat various other time
series, e.g., in social dynamics \cite{Cyc_weMyspace2012algorithm}, traffic on networks 
  \cite{Cyc_Traffic_dynamics2022}, the magnetisation fluctuations on the hysteresis loop \cite{mijatovic2022tuneable}, and other. 
More precisely, the time series having the length $T_{max}$ is divided
into overlapping segments, enumerated as $k=0,1,2,\cdots
k_{max}=T_{max}/m-1$, of the length $2m+1$, which overlap over
$m+1$ points. Then the polynomial fits $y^{(k)}(mk+\ell)$ over $\ell =0,1,2,\cdots 2m$ points in each segment are determined. Then the local trend $y_c(mk+i)$ over the overlapping points
is determined as 
$y_c(mk+i)= \frac{i}{m}y^{(k+1)}(mk+i) +
\frac{m-i}{m}y^{(k)}(m(k+1)+i)$,  balancing the contribution of the
polynomial in segment $k$ with the one of segment $k+1$; here
$i=0,1,2 \cdots m$ and $0<k<k_{max}$. In this way, the corresponding
polynomial contribution to the trend in the overlapped region decreases linearly with the distance from the segment's centre. Meanwhile, in the initial
$m+1$ points in $k=0$ and the final $m+1$ points in $k=k_{max}$
segments, the trend coincides with the actual polynomial fit.

\begin{figure*}[!htb]
\begin{tabular}{cc}                                                                                                       
\resizebox{32pc}{!}{\includegraphics{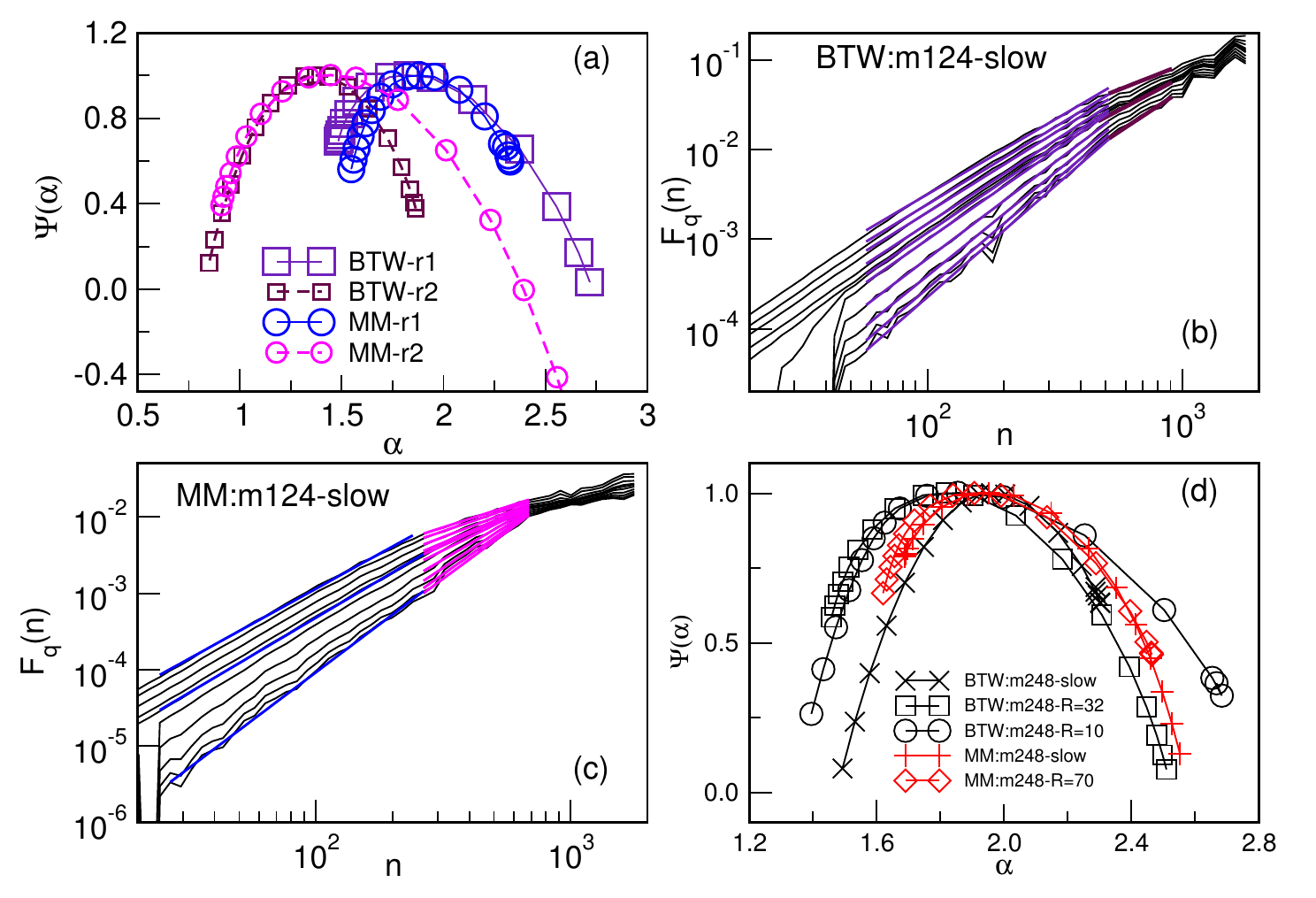}}\\                                                                                                          
\end{tabular}
\caption{Fluctuation function $F_q(n)$ vs time interval length $n$ for
the cyclical trend with $m=124$ in the stress fluctuation of 
BTW and MM sandpile automata with the adiabatic driving (b,c) and  their
singularity spectra  (a) corresponding to the region-1 (r1) and region-2
(r2) shown by the thick lines of the same colour on the $F_q(n)$
plots. (d) The singularity spectra for the large cycles ($m=248$)
for both SPA models with adiabatic (slow) and varied driving rates
$1/R$, values of $R$ are indicated in the legend; see text for details. }
\label{fig:Fqtrends2x}
\end{figure*}
For the studied case, the linear interpolation suffices,  and the
parameter $m$ is adapted, as stated above. 
As Fig.\ \ref{fig:ts2x} shows, these cyclical trends appear to have a lot
of harmonics, depending on the type of the SPA and the driving
rates. In contrast to social dynamics \cite{we_cycles_emo23}, where the primary cycles are introduced by the day-night fluctuations in the driving signal, the multiscale cycles in the SOC dynamics appear spontaneously. Thus, they can be visualised at different scales by adapting the parameter
$m$. For example, the red and yellow lines in Fig.\ \ref{fig:ts2x} correspond to $m=$124 and $m=$248, respectively. 
Visually, the emergent cycles in the case of BTW model with infinitely
slow driving is different from the ones found at finite driving rates
and processes in MM sandpiles  by all driving conditions. 
Moreover, a similarity between BTW at the finite driving rate and Manna
SPA is apparent, apart from the higher average value. 
In the following, the multifractal analysis and the related singularity spectra are determined to quantify these multiscale features of the identified cycles.

Applying the detrended multifractal  analysis
\cite{kantelhardt2002multifractal,pavlov2007multifractal,tadic2016multifractal} of time series, we
determine the generalised fluctuation function $F_q(n)$ as a function
of time intervals $n$ and determine
its scaling properties.  In this approach, the profile  $Y(i)
=\sum_{j=1}^i(C(j)-\langle C\rangle)$ of the series $\{C(j)\}$ is
constructed and  divide it in $N_s$ segments of the length $n$,
starting from the beginning and repeating from the end of the time
series$t=T_{max}$, which gives in total $2N_s=2Int(T_{max}/n)$ segments.
Then at each segment $\mu=1,2\cdots N_s$ the local trend $y_\mu(i)$ is
determined by polynomial fit and  the standard deviation around
it is computed  as $F^2(\mu,n) =
\frac{1}{n}\sum_{i=1}^n\left[Y((\mu-1)n+i)-y_\mu(i)\right]^2$ and
similarly  $F^2(\mu,n) = \frac{1}{n}\sum_{i=1}^n[Y(N-(\mu-N_s)n+i)-y_\mu(i)]^2$ for $\mu =N_s+1,\cdots 2N_s$.

The   fluctuation function $F_q(n)$ for the segment length $n\in[2,int(T_{max}/4)]$ is defined as
\begin{equation}
F_q(n)=\left(\frac{1}{2N_s}\sum_{\mu=1}^{2N_s}\left[F^2(\mu,n)\right]^{q/2}\right)^{1/q} \sim n^{H_q}  \ ,
\label{eq:fluctq}
\end{equation}
and computed for different positive and negative values of the
exponent $q\in [-4.5,4.5]$. The spectrum of the generalised Hurst exponent $H_q$ is extracted by fitting the power-law regions on
the lines for different $q$; in the case of monofractal, $H_q=H_2$ for
all $q$, where $H_2$ represents the standard Hurst exponent. 
Other multifractality measures \cite{kantelhardt2002multifractal} are readily determined from the spectrum $H_q$. In particular, the exponent  $\tau_q$ related to the standard (box probability) measure is given by $\tau_q=qH_q-1$. With the Legendre transform  $\Psi(\alpha)=q\alpha -\tau_q$, where $\alpha =d\tau/dq$ the  \textit{singularity spectrum} is obtained.  Thus, $\psi (\alpha)$ stands for a fractal dimension of the 
points having  the same singularity exponent $\alpha$, which indicates different power-law singularities  according to $\vert \nabla
r(j,\epsilon)\vert _{\epsilon \to 0}\sim \epsilon ^{\alpha (j)}$ at different
data points $j$ of the time series
\cite{pavlov2007multifractal,kantelhardt2002multifractal}.

\begin{figure}[!]
\begin{tabular}{cc}                                                                                                       
\resizebox{16pc}{!}{\includegraphics{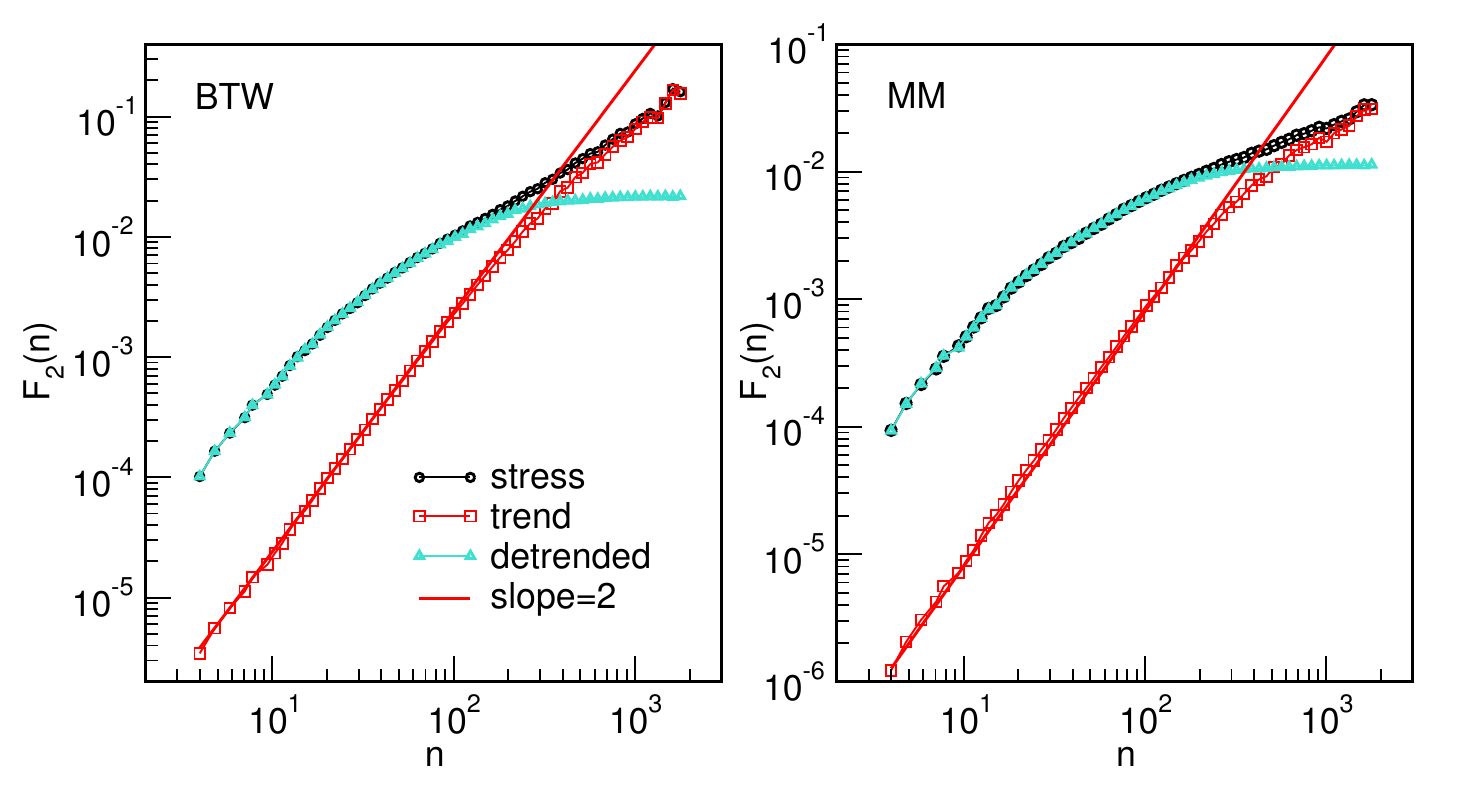}}\\
\resizebox{16pc}{!}{\includegraphics{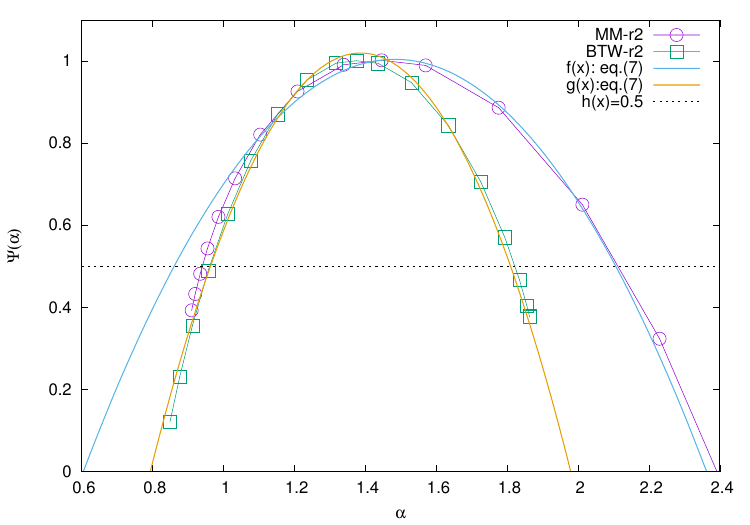}}\\                                                                                                             \end{tabular}
\caption{Top: Standard fluctuations for the stress, its cyclical
  trend and detrended signal in two SPA models, as indicated. Bottom:  The singularity spectra in region-2, fitted according to eq.\ (\ref{eq:psifit}), demonstrate quantitative differences in the dynamics of two SPA models. }
\label{fig:F2Psifits2x}
\end{figure}

In Fig.\ \ref{fig:Fqtrends2x}b,c, the fluctuation function $F_q(n)$ vs $n$ is
shown for the modulated cycles corresponding to the red lines 
at the slow-adiabatic driving rate both for the BTW and Manna SPA,
indicated in each panel. Interestingly, these fluctuation functions
exhibit two distinct regions, marked as region-1 (r1) and region-2
(r2), indicated by the straight lines of different colours. The
corresponding singularity spectra (symbols and lines with the matching
colours) are shown in Fig.\ \ref{fig:Fqtrends2x}a. In the region $r1$
with smaller time intervals,  the singularity spectra of both models
are centred around a similar value $\alpha_0\sim 1.94$, suggesting a
robust cyclical behaviour at these time scales. A broader spectrum
appears in the BTW case at both the small (right) and large (left side) fluctuations. For larger time scales, in the region-2 (r2 in the legend), however, the differences between two SPA appear to be more profound; see the discussion below and Fig.\ \ref{fig:F2Psifits2x}. Here, we show the results of a similar analysis of the fluctuation functions for the larger cycles ($m_2$, yellow lines in Fig.\ \ref{fig:ts2x}) and all
driving rates considered; the respective singularity spectra are shown
in Fig.\ \ref{fig:Fqtrends2x}d. As this Figure shows, the singularity spectrum for the BTW model in the case of adiabatic driving differs from the spectra at finite driving rates, both at large and small fluctuations. A
systematic broadening at the right side of the spectrum occurs with
the increased driving rate. Moreover, they show a relative similarity with the spectrum for the Manna SPA, where the lines for adiabatic and finite
driving rates practically coincide.

In the region $n<m$, the standard fluctuations around cyclical trends in two models show the trend's properties, i.e., $H_2=2$ and, similarly, the power spectrum with the exponent $\phi =2$ at high frequencies (not shown).
 Meanwhile, for much larger time intervals, the detrended signal saturates,
 and the trend and signal have similar fluctuations.

The shapes of the singularity spectra in region-2 demonstrate the essential differences in the dynamics of the two SPA
  models, even though they have similar avalanche exponents.
These differences can be quantified in analogy to spectra of damaged structures in Ref.\ 
\cite{damage2020multifractal},  by fitting the data with the expression
\begin{equation}
\Psi(\alpha)=A(\alpha -\alpha_0)^2 +B(\alpha -\alpha_0) +C
\label{eq:psifit}
\end{equation}
 and
introducing the  index
$I=\frac{2}{3}\frac{\alpha_0}{\Psi_{max}}w_{0.5}$, where $w_{0.5}$
stands for the width at the half of the $\Psi_{max}$. For example, the
data shown in Fig.\ \ref{fig:F2Psifits2x} bottom panel lead to
$I^{(BTW)}=0.743$ and $I^{(MM)}=1.031$, suggesting 
an increased amount of small  fluctuations in the stress of Manna SPA,
compared to BTW model.

\subsection{Sequences of dissipative events at finite driving
  rates\label{sec:avalanches}}
In this section, we focus on specific features of dissipative avalanches that are
responsible for the observed stress fluctuations. Recall that in the critical state of SPA,  the propagation of avalanches is a collective
dynamical process that does not change the stress unless an
avalanche hits the boundary. The amount of grains dissipated in such
events (outflow current $O_t$) can not be predicted; meanwhile, the sequences of such events 
contain information about the coherence of the collective dynamics
inside the sandpile.  

Considering the inter-event time related to the dissipative
avalanches with an outflow current that exceeds   
the lower bound of the dissipation taken into account, here  $O_t\ge 1 $.
Let $A_1, A_2,\ldots $ be consequent dissipative avalanches
such that the number of lost units of stress during these events is at least $1$ and $t_1$, $t_2$, $\ldots$ be the time of their occurrence.
Then we form the sample of inter-event intervals $\{t_2-t_1, t_3-t_2, \ldots\}$ and explore the frequencies $Z(\Delta t)$ of each observed
inter-event interval $\Delta t$.
\begin{figure}[!htb]
\begin{tabular}{cc}                                                                                                       
\resizebox{16pc}{!}{\includegraphics{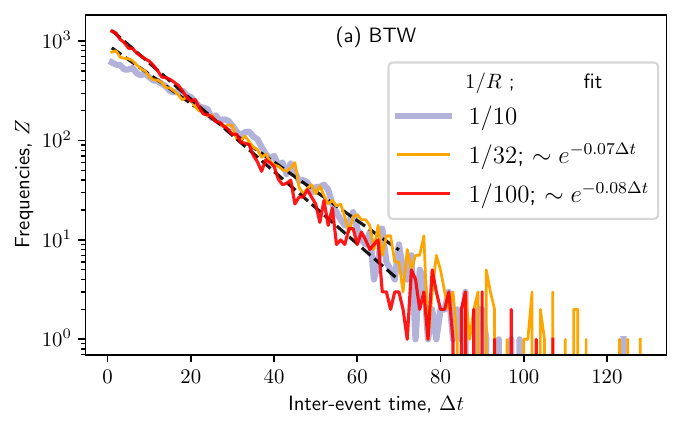}}\\
\resizebox{16pc}{!}{\includegraphics{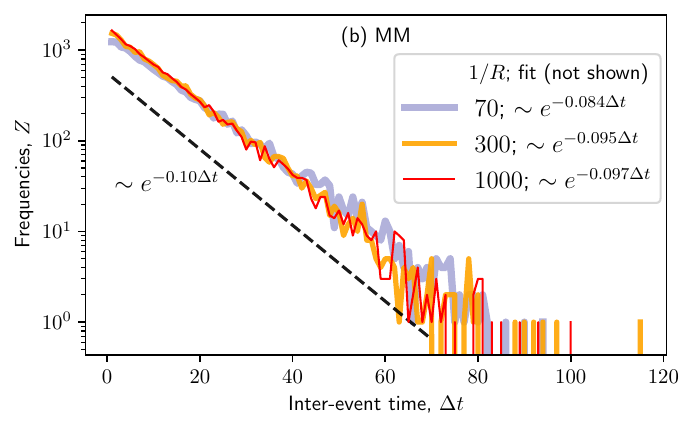}}\\
\end{tabular}
  \caption{ Frequencies of time intervals between two
    consecutive dissipative avalanches in the BTW (a) and MM (b) with 
    outflows  $\mathtt{O_t}\ge 1$.}
\label{f:interevent}
\end{figure}
Fig.~\ref{f:interevent} a,b visualize the corresponding frequencies. With the BTW models, the inter-event frequencies follow
the exponential function $e^{-\beta\Delta t}$ if the driving rate is sub-critical,
Fig.~\ref{f:interevent}a.
The values of $\beta$ slightly grow with $R$.
However, the exponential fit fails if the driving rate is super-critical.
Meanwhile, the exponential fit to inter-event distributions works for 
all values of the driving rate with the MM with slight variations in
$\beta$ as driving rate increases in the super-critical domain. These
findings are relevant for the events prediction; see Discussion.

\begin{figure}[!]
\begin{tabular}{cc}                                                                                                       
\resizebox{16pc}{!}{\includegraphics{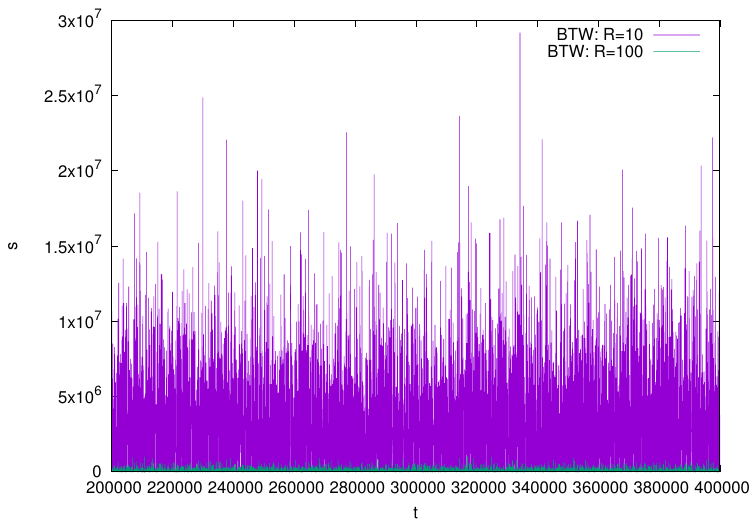}}\\
\resizebox{16pc}{!}{\includegraphics{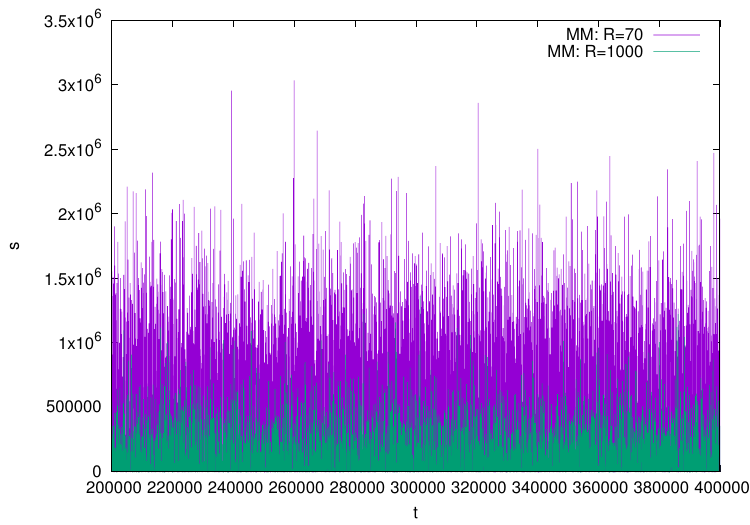}}\\                                    \end{tabular}
\caption{Sequences of the size $s$ of dissipative avalanches vs. time
  $t$ of all identified avalanches for two driving rates R
  indicated in the legend for BTW and MM sandpiles.  }
\label{fig:sedissip2x}
\end{figure}

\begin{figure*}[!htb]
\begin{tabular}{cc}                                                                                                       
\resizebox{24pc}{!}{\includegraphics{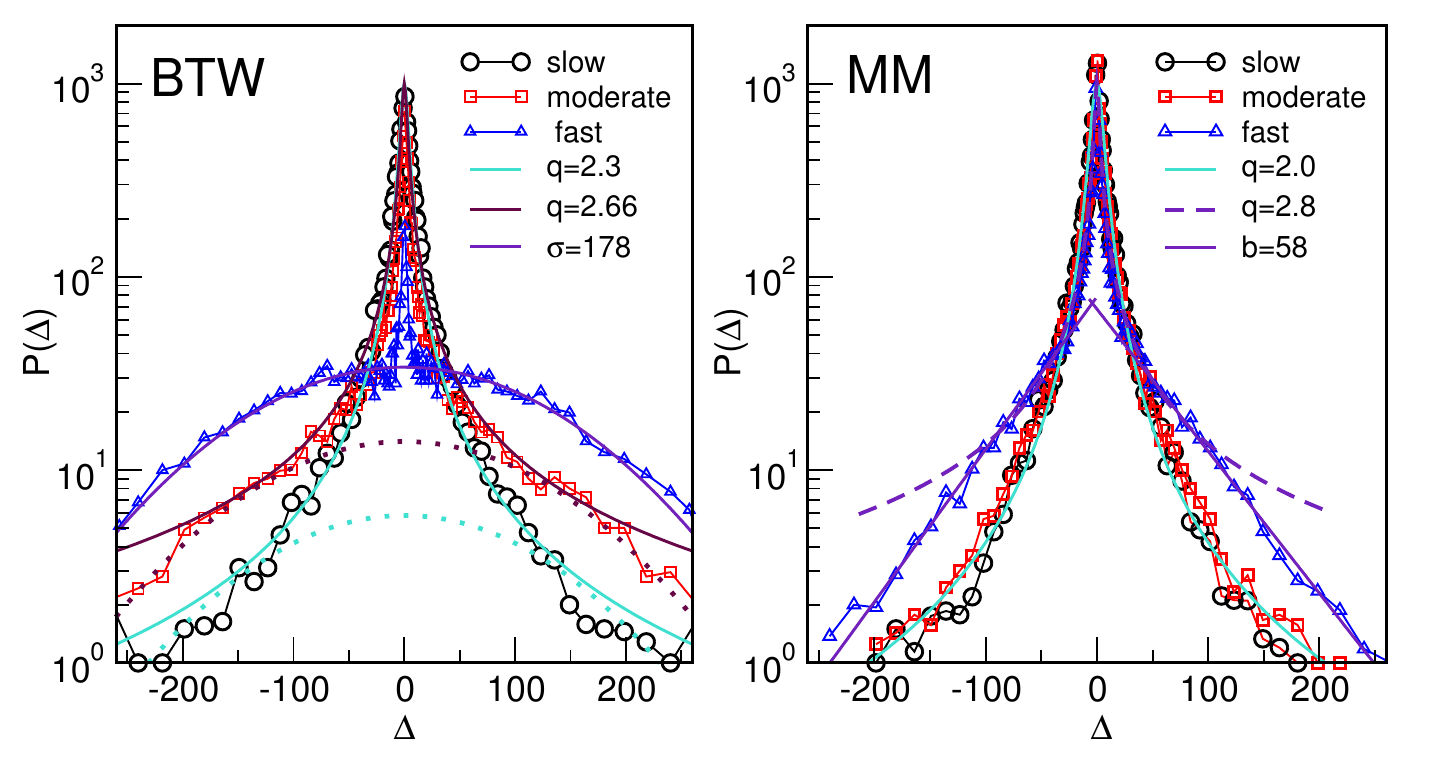}}\\
\end{tabular}
\caption{Distributions for the first returns $\Delta $ in the outflow current sequences of the dissipative avalanches for slow (adiabatic), moderate and fast
driving in BTW  and MM models. Legends show the fitted values of the pertinent parameters $q$ for q-Gaussian in Eq.(\ref{eq:qGauss}), $\sigma$--the width of the normal distribution, and $b$--for the exponential function fits; see text for more details.
}
\label{f:frstreturn}
\end{figure*}

The changes in the size of dissipative avalanches with the driving rate,
manifested in the avalanche distributions in Fig.\ \ref{f:sfrdiss}, is
illustrated in the sequence of dissipative avalanches in Fig.\
\ref{fig:sedissip2x}; cf.\ the stress fluctuations in Fig.\
\ref{fig:ts2x}. Two lines in each panel correspond to the smallest and largest
driving rates considered in the respective model.
As Fig.\ \ref{fig:sedissip2x} shows, dropping in the average additional 13 grains
per propagating avalanche induces more profound differences in the
sequences of the size of dissipative avalanches in BTW than in Manna models at similar conditions. These changes also indicate differences in the underlying dynamics that are further demonstrated in Fig.\ \ref{f:frstreturn} considering the distribution of first returns $\Delta = O_{t+1}-O_t$, the differences between the consecutive amounts in the sequences of outflow current.

Carried by the dissipative avalanches, the 1st-return distributions reflect the
level of coherence of the self-organised critical state.
Consequently, they are well-fitted by the Tsallis $q$-Gaussian
distribution 
\begin{equation}
P(\Delta )= A\left[1-(1-q)\left(\frac{\Delta}{\Delta_0}\right)^2\right]^{1/1-q} \ ,
\label{eq:qGauss}
\end{equation}
a generalisation of standard Gaussian distribution which arises in
the maximisation of Tsallis non-extensive entropy \cite{tsallis2009introduction}.
A comprehensive study in \cite{NEq_fTSexamples2014Pavlos} with
numerous real data examples show how $q$-Gaussian distribution arises in complex physical systems with fractal dynamics. 
The fitted values of the real-number parameter $q$, indicating the level of non-extensive dynamics with fractal features, are shown in the legend. 
Specifically, we find that in both models at slow (adiabatic) driving, the outflow first-returns can be described with the expression (\ref{eq:qGauss}) with $q\gtrsim 2$ within error bars $\pm 0.06$. (Note that the case $q=2$ corresponds to the Cauchy-Lorentz distribution.) 
However, when the driving rate is increased, only the central part
(corresponding to small differences $\Delta$) fits the expression (\ref{eq:qGauss}) with a larger $q$ value; meanwhile, the tails of the distributions follow a different law.  In particular, large returns $\Delta$ obey Gaussian distribution $f(\Delta)=Ae^{-(\Delta/\sigma)^2}$
with a different pre-factor $A$ and the width $\sigma \approx 178 \pm 5$ for all considered driving rates $R$ in BTW models, whereas the expression $g(\Delta)=Be^{-|\Delta|/b}$ fits the large-returns segments in the MM sandpiles at large driving rates.

\section{Discussion and Conclusions\label{sec:concl}}
We have studied the dynamics of sandpile automata with deterministic (BTW) and probabilistic (MM) rules on a square lattice with
variable driving rates, defined by the addition of grain at every $pR$ step during the avalanche propagation, where  $p=0,1,2,\cdots$, while $pR$ is smaller than the avalanche duration; $p=0$ corresponds to standardly considered adiabatic driving.
Having clearly distinguished slow time scale, defined by the sequence of
individual avalanches, from the fast time scale associated with the
intrinsic dynamics of avalanche propagation, we mainly focused on 
the stress fluctuations and the properties of the outflow current,
which maintains the sandpile's stationary state at every driving rate.
Our results revealed how some specific dynamical features of critical
sandpiles build up and are altered with increased driving rates. In particular:
\begin{itemize}
\item \textit{Cyclical trends} spontaneously appear in stress fluctuations at a slow time scale; collective dynamics of dissipative avalanches modulate these cycles such that they attain higher harmonics, described by the multifractal analysis. The singularity spectra are characteristic of the model dynamics and broaden with increasing driving rates.

\item \textit{Avalanches scaling loss } is demonstrated for the
  driving rates that exceed a certain limit, i.e.,  $1/R^{\star}$,
  depending on the model dynamic rules; 
  a more robust scaling behaviour is observed in MM than in BTW models. In the
  complementary parameter range, $<1/R^{\star}$, where additional
  grains are only sporadically dropped on the propagating avalanche,
  the scaling range, the exponents, and the finite size scaling properties might be altered.

\item \textit{Sequences of outflow current} induced by dissipative avalanches exhibit dramatic changes in the first-return distribution with the increased driving rate. A characteristic $q$-Gaussian, with $q\gtrsim 2$ for both models at slow driving,  gradually changes towards Gaussian distribution in BTW sandpiles. In contrast, an exponential distribution applies to the significant first returns in the MM case at large driving rates. 
\end{itemize}

A number of open questions remains for future study. In particular,
regarding the scaling loss and the existence of a finite critical driving rate
$1/R^{\star}$,  a rigorous finite-size scaling analysis with a correct identification
of scaling variables is necessary, in analogy to seminal studies of
nonequilibrium disordered ferromagnets with avalanching dynamics \cite{BHN_FSStheoryplusPRB2003RecheVives,BHN_numericalRcPRL2011Djole}.
In complex dynamical systems, the $q$-Gaussian distributions are often detected with coherent dynamics of aggregates related to many interdependent components. The sandpile dynamics is characterized by dissipative avalanches that expand at different parts of the system, thus contributing to the oscillations of the average stress and rare huge drops in stress that generate global dependencies. Interestingly, the significant driving rate 
eliminates these interdependences, ruling out the $ q-$ Gaussian
features.
Moreover, the exponential inter-event distribution indicates that the sequence of
events itself is unpredictable, i.e., new events cannot be predicted
based on their history only. Therefore, if dissipative extremes in
the MM with a large driving rate inherited the
predictability known (see \cite{SOC_rainfalls2015,
  shapoval2022universal}) for the original Manna model, an
efficient prediction algorithm must be built on specific scenarios
preceding the events-to-predict.
On the contrary, unpredictability of the BTW sandpile may be turned to an efficient
prediction of extremes with the models at  large driving rates
because yet the sequence of these events itself exhibits the traces
of predictability related to the observed non-exponential
distribution. 
The corresponding algorithms are discussed in \cite{molchan1997, shapoval2010prediction}.

In summary, we highlight that the above-described
dynamical features are different in these studied sandpile models
despite their well-known similarity regarding the scaling exponents of
avalanches at adiabatic driving. These findings indicate that the fundamental principles of SOC---building the collective behaviours from
specific microscopic dynamic rules---also apply to the pathways towards losing
criticality when the driving conditions are changed.   The robust
appearance of cycles, even at significant driving rates, indicates that the
collective fluctuations at different scales exist beyond the scaling
of avalanches. Therefore, studies of cyclical trends in time series of
a relevant variable may be utilised as a critical signature of
self-organisation in the underlying dynamics of many complex systems
when interactions and driving forces are less apparent.

\section*{Acknowledgments}

B.T.  acknowledges the financial support from the Slovenian
Research Agency under the program P1-0044. A.S. is thankful to A. Orpel and A. Nowakowski for the discussion of the paper.


\section*{Statements}

Data sharing is not applicable as no new data is generated. 

The authors declare the absence of the conflict of interest.

\end{document}